\documentclass[preprint,showpacs,preprintnumbers,amsmath,amssymb]{revtex4}

\usepackage{graphicx}

\begin{document}
\title {In-medium NN cross sections determined from stopping and
collective flow in intermediate-energy heavy-ion collisions}

\author {Yingxun Zhang$^{1}$
\email[1]{zhyx@ciae.ac.cn}, Zhuxia Li$^{1,2,3)}$
\email[]{lizwux@ciae.ac.cn}, Pawel Danielewicz$^{4)}$}
\address{
 1) China Institute of Atomic Energy, P. O. Box 275 (18),
Beijing 102413, P. R. China\\
2) Center of Theoretical Nuclear Physics, National Laboratory of
Lanzhou Heavy Ion Accelerator,
 Lanzhou 730000, P. R. China\\
 3) Institute of Theoretical Physics, Chinese Academic of Science,
Beijing 100080\\
4) National Superconducting Cyclotron Laboratory, Michigan State
University, East Lansing, Michigan 48824, USA}
\begin{abstract}
In-medium nucleon-nucleon scattering cross sections are explored
by comparing results of quantum molecular dynamics simulations
to data on stopping and on elliptic and directed flow in intermediate-energy
heavy-ion collisions.  The comparison points to in-medium cross sections
which are suppressed at low energies but not at higher energies.
Positive correlations are found between the degree of stopping
and the magnitudes of elliptic and directed flows.
\\
\end{abstract}
 \maketitle

One of the main goals of research in the area of heavy ion
collisions (HICs) at intermediate energies has been the
determination of bulk properties of nuclear matter, such as the
nuclear equation of state (EoS).  While a considerable
progress~\cite{Dan02, Fuch06, Demo90, Gutb90, Reis97} including
that in subthreshold kaon production which provides complementary
information on the EOS\cite{kaon} has been reached in determining
the EoS at supranormal densities, relying on reaction data, the
uncertainties are still very large. To access the EoS, it is
necessary to describe reaction observables \cite{Demo90, Gutb90,
Reis97}, such as those quantifying the collective motion of
nuclear matter, within reaction theory \cite{Dan02, Fuch06,
Stoc86, Gale90, Pan93, LiBA99, Zhou94}. The transport models
employed in the description of central reactions have included the
quantum molecular dynamics approaches in its QMD \cite{Aich86,
Hart89} and ImQMD \cite{Wang2004, Zh05, Zh06} (with Im for
Improved) variants, as well as the Boltzmann-Uehling-Uhlenbeck
(BUU) approaches~\cite{Bert84, Aich85, QLi89}. The two main
ingredients of the nuclear transport are the nucleonic mean fields
and nucleon-nucleon binary scattering cross sections (NNCS). The
employed cross sections affect virtually any observable from a
central reactions and constraining those cross sections is
essential for reducing the EoS uncertainties \cite{Pan93}. Also
the in-medium cross-sections are of interest for their own sake,
as they underly the viscosity and other nuclear transport
coefficients \cite{Shi03}.  Best for constraining the cross
sections are the stopping observables \cite{Reis04, Pers2002}
 that reflect the cross sections in the most direct manner.  However, the correlation
\cite{Reis04} with flow observables, such as used for determining the EoS, is also
of interest.  In this paper, for constraining the cross sections, we shall employ the ImQMD model
\cite{Wang2004, Zh05, Zh06}.

Principally, within theory the mean field and cross sections of
transport calculations should be linked to the same microscopic
nucleon-nucleon interactions.  In practice, this has turned out to
be overly ambitious resulting in the phenomenological strategies
in transport of varying independently the mean field, related to
EoS, and the NNCS.  Microscopically, far fewer calculations have
been done of the in-medium cross-sections than of the EoS and of
the mean fields. Specifically, the zero-temperature in-medium
elastic NNCS (ENNCS) have been studied within the relativistic
\cite{Haar87, LiGQ93, Fuchs01} and non-relativistic Brueckner
approaches~\cite{Schul97, Kohn98}.  In Refs.\ \cite{LiGQ93,
Fuchs01}, employing the Brueckner relativistic approach, it was
found that, compared to the free ENNCS, the in-medium ENNCS were
suppressed at low relative momenta and less suppressed at higher
relative momenta and even enhanced slightly depending on the
medium density. The in-medium ENNCS have also been studied by
using the closed time path Green's function (CTPGF) approach
\cite{Mao94, Dick99, QFLi2000, QFLi2004} employing, in particular,
the QHD-I and QHD-II effective Lagrangians, with the mean field
and in-medium ENNCS derived self-consistently for the same
effective interaction. The latter studies  \cite{Mao94, QFLi2000,
QFLi2004} produced in-medium correction factors which were
different for the cross sections of like,
$\sigma^{*}_\text{nn,pp}$ and unlike, $\sigma^{*}_\text{np}$,
nucleons.  Otherwise, the zero-temperature ENNCS from the
Dirac-Brueckner \cite{LiGQ93, Fuchs01} and CTPGF \cite{Mao94,
QFLi2000, QFLi2004} have exhibited similar qualitative features
but the results have differed in the details of their dependance
on density and energy.  The temperature dependence of the
in-medium ENNCS has been investigated in \cite{QFLi2004} within
the CTPGF approach and a general increase in the ENNCS with
temperature was found.  Elsewhere~\cite{Dick98}, though,
difficulties have been pointed out in defining the cross sections
within a nuclear medium.  Those difficulties might relegate the
cross sections utilized in transport to strictly phenomenological
quantities. In the comparisons of nuclear reaction simulations to
data, indeed, most often phenomenological cross section
parameterizations have been employed, as e.g.\ represented by the
formula
\begin{equation}
\sigma_\text{NN}^{*}=(1-\eta\rho/\rho_{0})\sigma_\text{NN}^{free}
\label{eq:sigma}
\end{equation}
where $\eta=0.2$ \cite{Klak93, Pers2002, Dani2000, Fuchs01}.  Also, the
in-medium ENNCS scaled by the effective mass,
$\sigma_\text{NN}^{*}/\sigma_\text{NN}^\text{free}=(m^{*}(\rho,p)/m)^2$ has also
been employed in comparisons \cite{LI05} to data from heavy-ion collisions (HIC).
The latter scaling presumes that, for given relative momenta, the matrix elements of interaction are not changed
between the free space and medium.

The purpose of this work is to draw conclusions on the in-medium ENNCS using
recent data on stopping and elliptic and directed flows, obtained with a good centrality
selection, from
collisions of Au + Au and other symmetric or near-symmetric systems \cite{Reis04,Luka2005,Andr05}.
For those data, a high degree of correlation was found
between the degree of stopping and the strength of collective flow.
In our investigations we rely on the recent version ImQMD05 of
the ImQMD model \cite{Wang2004,Zh05,Zh06}.

Within the ImQMD05 model, the mean fields acting on nucleon wavepackets are derived
from an energy functional where the potential energy $U$ includes the full Skyrme potential energy with just
the spin-orbit term omitted:
\begin{equation}
U=U_\rho + U_\text{md}+U_\text{Coul} \, .
\end{equation}
Here, $U_\text{Coul}$ is the Coulomb energy, while the nuclear contributions can be represented
in local form with $U_\text{$\rho$,md}= \int \text{d}^3 r \, u_\text{$\rho$,md}$ and
\begin{eqnarray}
u_{\rho}&=&\frac{\alpha }{2} \, \frac{\rho ^{2}}{\rho _{0}}+\frac{\beta }{\gamma +1} \,
\frac{\rho ^{\gamma +1}}{\rho _{0}^{\gamma }}+\frac{g_\text{sur}}{2\rho _{0}} \,
(\nabla \rho
)^{2}
+ \frac{g_\text{sur,iso}}{\rho_{0}} \, [\nabla(\rho_\text{n}-\rho_\text{p})]^{2}
\nonumber\\
&&
+(A\rho^{2}+B\rho^{\gamma+1}+C\rho^{8/3}) \, \delta^{2}
+g_{\rho\tau} \, \frac{\rho^{8/3}}{\rho_{0}^{5/3}} \, , \label{13}
\end{eqnarray}
where
$\delta=(\rho_\text{n}-\rho_\text{p})/(\rho_\text{n}+\rho_\text{p})$,
$\rho=\rho_\text{n}+\rho_\text{p}$ and $\rho_\text{n}$ and $\rho_\text{p}$ are the neutron
and proton densities, respectively.  The energy associated with the mean-field momentum
dependence may be represented as
\begin{equation}\label{eq:modep}
u_\text{md} = \frac{1}{2 \rho_0} \sum_{N_1,N_2= \text{n,p}}
\frac{1}{16 \pi^6} \int \text{d}^3 p_1 \, \text{d}^3 p_2 \,
 f_{N_1}({\bf p}_1) \, f_{N_2}({\bf p}_2)\,
 1.57 \, \left[ \ln{\left( 1. + 5. \times 10^{-4} (\Delta p)^2 \right)} \right]^2 \, ,
\end{equation}
where $f_N$ are nucleon Wigner functions, $\Delta p = |{\bf p}_1 - {\bf p}_2|$, the energy is in MeV
and momenta are in MeV/c;
the resulting interaction between wavepackets is such as in Ref.~\cite{Aich87}.
The coefficients utilized in (\ref{13}) can be
transcribed, see \cite{Zh06}, onto those usually specified for the Skyrme interactions.
In this work,
the SkP and SLy7 Skyrme interactions are employed.  Both of those interactions give rise to
an incompressibilty of $K \sim 200$~MeV and produce an EOS consistent with the features of
collective flow in HICs from Fermi to relativistic energies.  However, the symmetry energies
associated with those two interactions are different.  It should be stressed that the expression
for the energy (\ref{13}) is more complete here than in the preceding Ref.~\cite{Zh06}.

Within this paper, three different phenomenological forms of
in-medium ENNCS are utilized. The first set
$\sigma_\text{NN}^{*(1)}$ are the cross sections given by
Eq.~(\ref{eq:sigma}) with $\eta=0.2$ and with the free cross
sections described in terms of the parameterization of
Ref.~\cite{Cug96}. The second set $\sigma_\text{NN}^{*(2)}$ are
cross sections calculated within the CTPGF approach of
Ref.~\cite{QFLi2000}, following the QHD-II Lagrangian.  For
calculational convenience, within ImQMD the formula
(\ref{eq:sigma}) is, though, employed with $\eta$ made dependent
on density and energy and its dependence fitted to the CTPGF
results. For reference, in Fig.~1 we present both the cross
sections, $\sigma_\text{np}^{*(2)}$ and $\sigma_{nn,pp}^{*(2)}$ in
panels (a) and (b), and the suppression parameters,
$\eta_\text{np}^{(2)}$ and $\eta_{nn(pp)}^{(2)}$ in panels (c) and
(d), as a function of c.m.\ energy $\sqrt{s}$ at different
densities.  One can see that $\sigma_\text{np}^{*(2)}$ changes
little with density and is nearly the same as in free space.  On
the other hand, the cross section $\sigma_\text{nn,pp}^{*(2)}$
tends to be suppressed at lower energies, $\sqrt{s} \lesssim
2.05\, \text{GeV}$ and enhanced at higher.  Differences in the
features of the two cross sections are associated with the
differences between the $T=0$ and the $T=1$ channels and, in
particular, presence of a low-energy resonance in the $T=1$
channel and effects of $\rho$ exchange.
 The third set constitutes and {\em ad hoc} parameterization, inspired by
the CTPGF results, aiming at the description of the excitation
function for elliptic flow in the midrapidity region of
$|y_\text{cm}/y_\text{cm}^\text{beam}|<0.1$ in Au + Au
collisions~\cite{Zh06}. In that parameterization, the common
cross-section modification parameter, $\eta^{(3)}=
\eta_\text{np}^{(3)} = \eta_\text{np}^{(3)}$, depends on the beam
energy for reaction, $E_\text{beam}$, in the following manner:
$\eta^{(3)} =0.2$ for $E_\text{beam} < 150 \, \text{AMeV}$,
$\eta^{(3)} =0$  for $150 \, \text{AMeV} < E_\text{beam} < 200 \,
\text{AMeV}$, $\eta^{(3)} = -0.2$  for $200 \, \text{AMeV} <
E_\text{beam} < 400 \, \text{AMeV}$, and $\eta^{(3)} = -0.4$  for
$400 \, \text{AMeV} < E_\text{beam}$. In this paper, we actually
confine ourselves to the HIC energy range of $ E_\text{beam} < 400
\, \text{AMeV}$ since, on one hand, our model is nonrelativistic
and, on the other, higher energies require a consideration of the
inelastic cross section that we are not prepared to carry out at
this moment.

First, we investigate the impact of in-medium NNCS onto the model predictions for elliptic and directed
flows in Au + Au collisions.  Excitation functions for both flow observables have been
determined experimentally \cite{Luka2005,Reis04,Andr05}.  The top panels in Fig.~2 show the excitation
function of the elliptic flow parameter $v_2$ for $Z \le 2$ particles in the midrapidity region
of $|y_\text{cm}/y_\text{cm}^\text{beam}|<0.1$.  The $v_2$ values have been obtained using the rotated
frame \cite{Luka2005}.  The bottom panels in Fig.~2 show the excitation function
of maximal scaled directed flow in Au + Au collisions at $b \simeq 5 \, \text{fm}$ ($b/b_\text{max} \sim 0.38$).
The scaled flow is
defined with~\cite{Reis04}
$p_{x\text{dir}}^{(0)} = p_{x\text{dir}}/u_\text{cm}^\text{beam}$, where
$u_\text{cm}^\text{beam} = \gamma_\text{cm}^\text{beam} \beta_\text{cm}^\text{beam}$ and
$p_{x\text{dir}} = \sum \text{sgn}(y) \, Z \, u_{x} / \sum Z$.  The sums in the directed flow
extend over all particles with charge number $Z \le 10$.  The data and calculations are represented in
Fig.~2 with closed and open symbols, respectively.  Lines generally merely guide the eye,
 except for the dashed lines in the bottom panels.  In these cases, the dashed lines illustrate the
 change in the directed flow data after applying the correction for reaction-plane fluctuation.
The calculations in the left panels of Fig.~2 have been carried out with the SkP and SLy7 mean fields
and $\sigma^{*(1)}$ cross sections.  The differences between the calculated $v_2$ values
in the panel (a) of Fig.~2 are limited and can be due to the difference in the symmetry energy
between the two mean-field models.  Notably, there are some differences between the data sets
in the figure too.  One more significant difference which develops between the calculations
for different mean fields in Fig.~2 is in
$p_{x\text{dir}}^{(0)}$ in panel~(c) at
$E_\text{beam} > 150 \, \text{AMeV}$.  The calculations in the right panels of Fig.~2 are for the SkP mean
field in combination with
the three different cross-section models.  The sensitivity of flow to the cross sections is similar to the sensitivity
to the mean-field models.  It is limited on absolute scale but gets more enhanced at higher energies.
The deviations between the models
are similar to the deviations of models from data.  From the presented
mean-field and cross-section combinations, the SkP mean-field in combination with $\sigma^{*(3)}$ cross sections describe
the flow data best.

We next turn to the impact of NNCS on nuclear stopping.  Recently, in the experiment~\cite{Reis04}, the ratio of the
rapidity variance in the transverse direction to the rapidity variance in the longitudinal direction,
$vartl$, has been used a measure of the nuclear stopping.  The longitudinal rapidity for the ratio
is defined in the standard manner in the cm system.  The transverse rapidity is defined by replacing
the longitudinal direction in the definition with a random transverse direction.
Figure~3 shows a variety of results pertinent to $vartl$.  The results are for charged particles
with $Z=1-6$ from central collisions of symmetric or near-symmetric systems,
with the contributions of different particles weighted with~$Z$.
The panels (a) and (b) in Fig.~3 show calculated distributions in longitudinal and transverse rapidities
in 400 AMeV Au + Au
collisions at $b/b_\text{max}<0.15$.
The corresponding $vartl$ values for different calculations are quoted in those panels.
The panel (c) compares chosen calculated Au + Au excitation functions for $vartl$
to data.  Finally, the panel (d) compares the calculated dependence of $vartl$ on system charge to data at 400 AMeV.

From the different panels of Fig.~3, the panel (a) tests the sensitivity of the calculated rapidity distributions to a
utilized mean field.  It is apparent that that sensitivity is quite weak.  Quantitatively,
when switching from the SkP to Sly7 mean field, $vartl$ increases just by 0.015 i.e.\ relative 2\%.
The panel (b) next tests the sensitivity of the calculated rapidity distributions to a
utilized cross section.  The sensitivity is much greater here, with $vartl$ changing by 0.22 or
32\% when switching from the $\sigma^{*(1)}$ to $\sigma^{*(3)}$ cross section.
Thus, there is a good chance to restrict the in-medium cross sections using measured $vartl$ but less
chance to restrict the mean field.  What is important in Fig.~3(b) is that the larger the cross
section in the given energy region the more similar are the transverse and longitudinal
distributions and the larger the corresponding $vartl$ value.  The panels (c) and (d) compare
to data the results obtained with the SkP mean field and two of the in-medium cross sections,
$\sigma^{*(1)}$ and $\sigma^{*(3)}$.  At lower energies a semiquantitative agreement with the Au + Au
data is found and there the cross sections coincide.  However, at higher energies
the Au + Au data favor $\sigma^{*(3)}$ NNCS.  The
 $\sigma^{*(1)}$ NNCS appears excessively reduced and this also concerns $\sigma^{*(2)}$ given
 the 400 AMeV $vartl$ value of 0.805 in the panel (b).  Looking next at the system charge dependence
 of $vartl$ in the panel (d), it appears that an optimal in-medium cross at 400~AMeV could be a bit
 lower than $\sigma^{*(3)}$.  Of the three cross sections, nonetheless, $\sigma^{*(3)}$ compares best with
 the data.

Experimentally, the values of $vartl$ maximize at $E_\text{beam}
\sim 400 \, \text{AMeV}$ and so do the values of
$p_{x\text{dir}}^{(0)}$.  At the same time, the elliptic flow
values minimize in this energy region \cite{Luka2005}.  If the
values of $vartl$ at different energies for the Au + Au system are
considered simultaneously with the values of maximal scaled
directed flow, a positive relatively narrow correlation is found
between the observables \cite{Reis04}. Importantly, that
correlation is between observables determined, on one hand, in the
most central $b/b_{max}<0.15$ collisions and, on the other, the
semicentral $b/b_{max}\sim 0.38$ collisions.  This suggests, from
the experimental side, that both observables probe bulk nuclear
properties which come into play at specific densities and
excitations reached at a particular beam energy. Between the $vartl$ and $v_{2}$
values, again determined at different centralities, a negative correlation is
expected.
Interestingly, when other symmetric or near-symmetric systems are
considered, and specifically Ni + Ni, Ru + Ru and Xe + Cs, the
resulting correlations appear to extend the correlation lines
found for Au + Au, see Fig.~4, with the left panel (a) showing the
$p_{x\text{dir}}^{(0)}$-$vartl$ correlation and the right panel
(b) the $v_2$-$vartl$ correlation.  The system systematics appears
further to underscore, from the experimental side, that bulk
nuclear properties get tested by the observables.

It is obviously important to find out whether the ImQMD05 model
produces correlations comparable to those measured and whether the
calculated and measured correlations agree quantitatively.
Fig.4(a)and (b) show the correlation of maximal $p_{xdir}^{(0)}$
and $v_{2}$ with $vartl$, respectively. All of these results are
for the reaction systems $Ni+Ni, Ru+Ru, Xe+Cs, Au+Au$ at
$E_{beam}=250, 400AMeV$. While the data are represented with
crosses and stars in Fig.~4, the calculations are represented with
planar-figure symbols.  For each of the represented cross
sections, i.e.\ either $\sigma^{*(1)}$ represented by open
symbols, or $\sigma^{*(3)}$ represented by closed symbols, in
combination with the SkP mean field, the calculations produce
approximate correlation lines in the planes of flow vs $vartl$,
with positive slope in the case of $p_{x\text{dir}}^{(0)}$ and
negative in the case of $v_2$. One of the calculated correlations,
$v_2$-$vartl$ for $\sigma^{*(1)}$, is somewhat broader than the
others.  To aid the eye, the correlation areas for $\sigma^{*(1)}$
are additionally marked with shadowing. Within the reaction
simulations, the correlations between stopping and flow may be
primarily associated with the collisions governed by cross
sections.  The greater the momentum transfer in collisions the
higher the stopping.  However, the occurring local randomization
due to collisions also enhances hydrodynamic behavior and, thus,
enhances the magnitude of flow observables. From the calculated
correlations in Fig.~4, only the two obtained using
$\sigma^{*(3)}$ reasonably agree with data.  Notably, with the
stopping and flow observables simultaneously growing in magnitude
with collision number, the correlation lines associated with the
different cross sections could have actually coincided. However,
$vartl$ and flow observables depend on interactions differently.
The $vartl$ observable exhibits a strong dependence on cross
sections and little on mean field, while flow observables depend
to a comparable extent on cross sections and mean field. As a
result, the correlation lines can end up being different for the
different cross sections.

In summary, within the ImQMD05 model we have investigated the
impact of in-medium NNCS and mean fields on stopping and flow in
HICs at intermediate energies.  The $vartl$ observable,
quantifying the stopping, depends strongly on NNCS and little on
mean-field details.  On the other hand, the $v_2$ and
$p_{x\text{dir}}^{(0)}$ flow observables depend comparably on
uncertainties in the NNCS and mean fields.  Within the upper range
of the energies we have investigated, both the common NNCS
parameterization represented by $\sigma^{*(1)}$ and the CTPGF
cross section represented by  $\sigma^{*(2)}$ appeared excessively
reduced in the medium to produce observed $vartl$ values
\cite{Reis04}. Also the magnitudes of flow observables calculated
for those cross sections turned out to be low compared to the
experiment.  The agreement between the data and calculations gets
much improved when employing an {\em ad hoc} cross section
parameterization guided by CTPGF and represented by
$\sigma^{*(3)}$.  For that parameterization the nucleon-nucleon
cross sections in the medium are suppressed at low energies and
enhanced at higher energies, i.e. the extracted in-medium
nucleon-nucleon cross sections depend on both density and energy.
 In this respect it is also in agreement with the prediction of
\cite{LiGQ93,Fuchs01}. The results clearly do not represent the
last word on in-medium cross sections as medium-variation of the
cross sections in the parameterization has been clearly
oversimplified. From the side of observables, those that we
examined do not strongly differentiate between the like and unlike
nucleon cross sections.

\begin{center}
{\bf Acknowledgments}
\end{center}
 This work was supported by the National Natural
Science Foundation of China under Grant Nos.\ 10175093, 10675172 and 10235030,
by the Major State Basic Research Development Program under Contract
No.\ G20000774 and by the U.S.\ National Science Foundation, Grant No.\ PHY-0555893.

\newpage
\begin{description}
\item[\texttt{Fig.1}] Energy dependence of in-medium cross-sections
$\sigma_\text{np}^{*(2)}$ and
$\sigma_\text{nn,pp}^{*(2)}$, in panels (a) and (b), and of the suppression parameters
$\eta_\text{np}^{(2)}$
and $\eta_\text{nn,pp}^{(2)}$,
in panels (c) and (d), at selected densities.  The cross sections have been obtained within
the CTPGF approach with
QHD-II effective lagrangian.

 \item[\texttt{Fig.2}]  Flow excitation functions in Au + Au collisions.  The top panels display
 midrapidity elliptic flow.  The bottom panels display maximal scaled directed flow.  The data
 are represented by solid symbols and calculations by open symbols.  Lines generally merely guide the eye,
 except for the dashed lines in the bottom panels.  In these cases, the dashed lines illustrate the
 change in the directed flow data after applying the correction for reaction-plane fluctuation.
The left panels in the figure illustrate
 the sensitivity of calculations to the employed mean field.  The right panels illustrate the sensitivity
 of calculations to in-medium cross sections.

\item[\texttt{Fig.3}]
Comparison of characteristics of central colliding systems in the longitudinal and transverse rapidities.
Panels (a) and (b) display the calculated rapidity distributions of particles from central
($b/b_\text{max}<0.15$) 400 AMeV
Au + Au collisions and illustrate, respectively, the sensitivity of calculations to the mean field and
to the cross sections.  The rapidities are normalized to the beam cm rapidity.  Panels (c) and (d)
display the calculated and measured dependencies of the variance ratio $vartl$, respectively, on energy for the Au + Au
collisions and on net system charge at collision energy 400 AMeV.  The data are from the FOPI Collaboration
\cite{Reis04}.  The calculations have been done for the SkP mean field combined with either
$\sigma^{*(1)}$ or $\sigma^{*(3)}$ cross sections.

\item[\texttt{Fig.4}]  Correlation between $vartl$ and flow
observables: $p_{x\text{dir}}^{(0)}$ in the panel (a) and $v_2$ in
(b).  Both the data \cite{Reis04,Luka2005} and calculations are
represented.

\end{description}

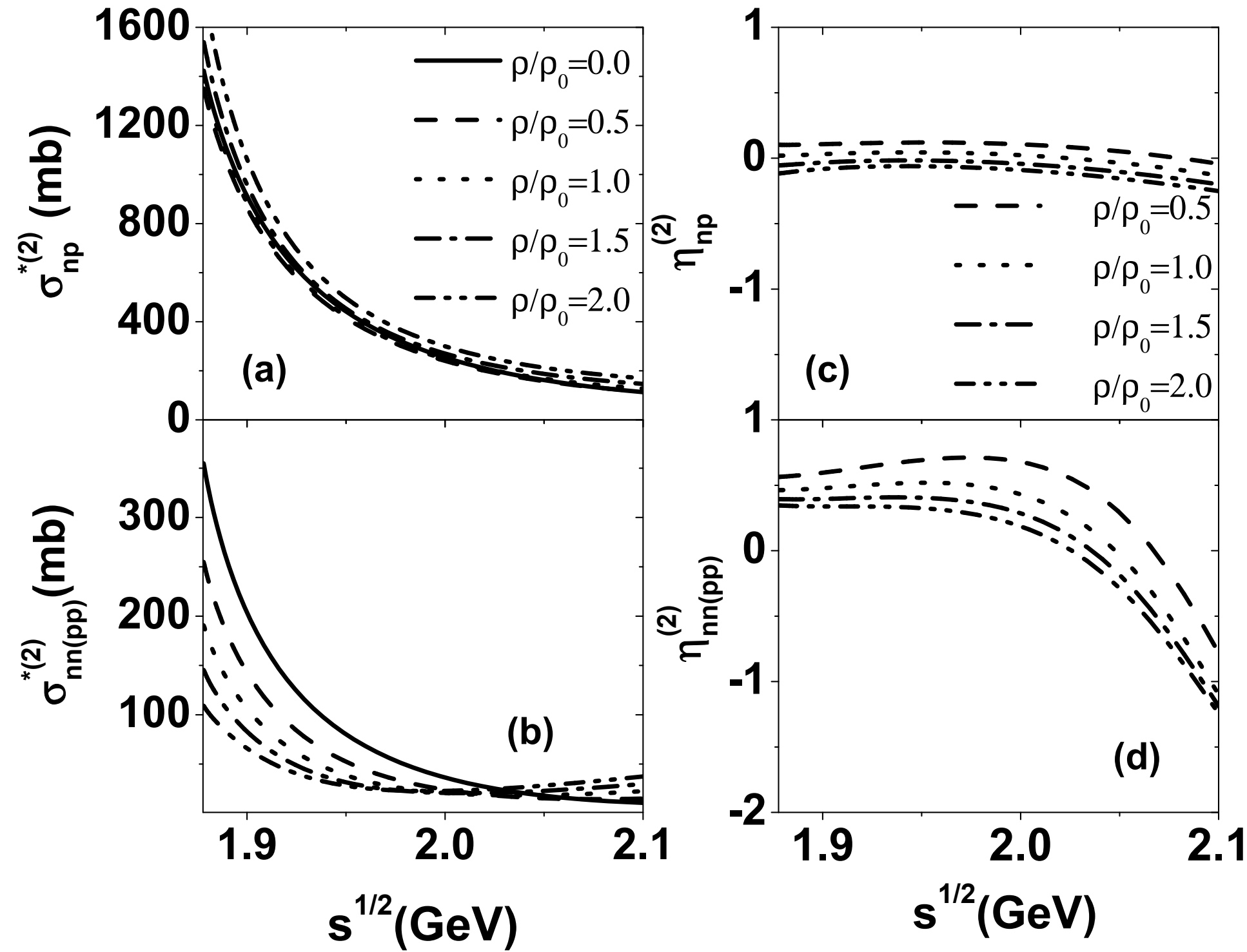
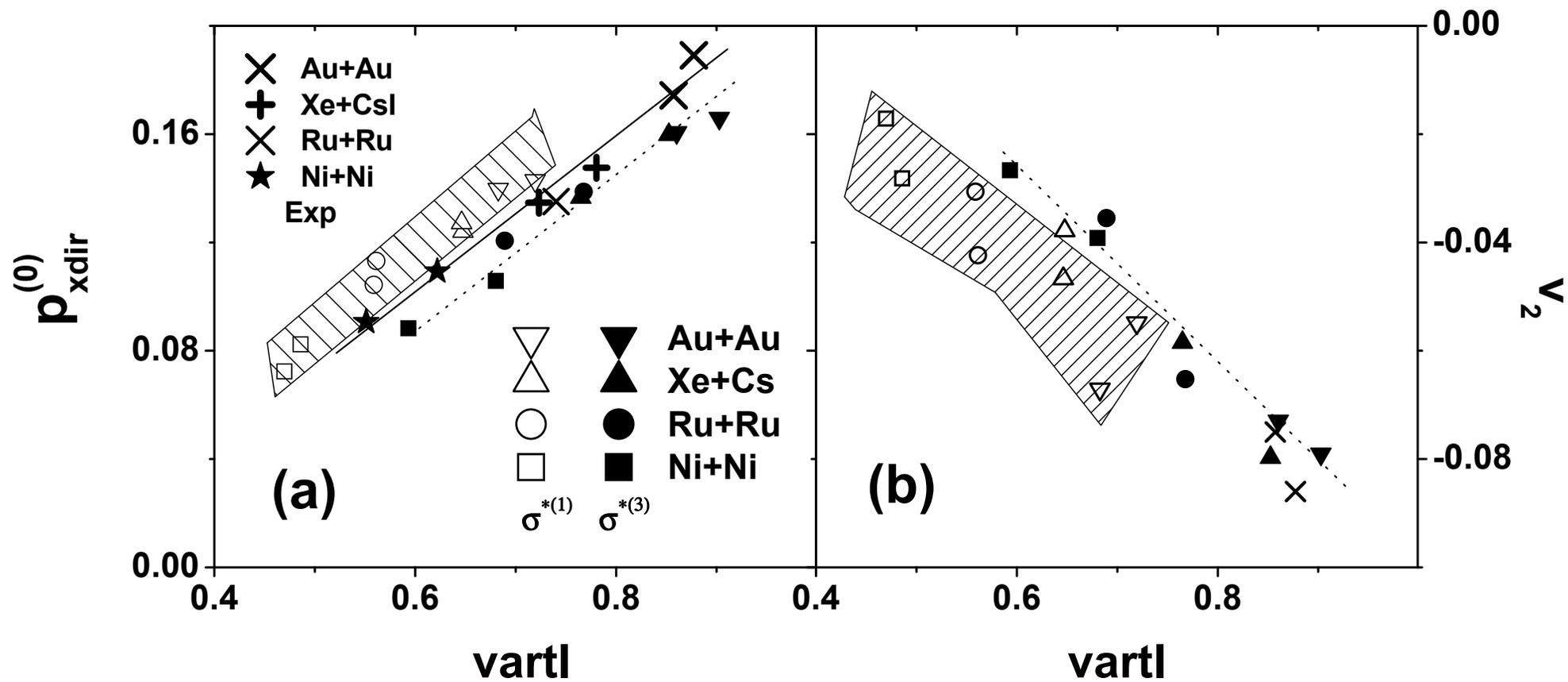
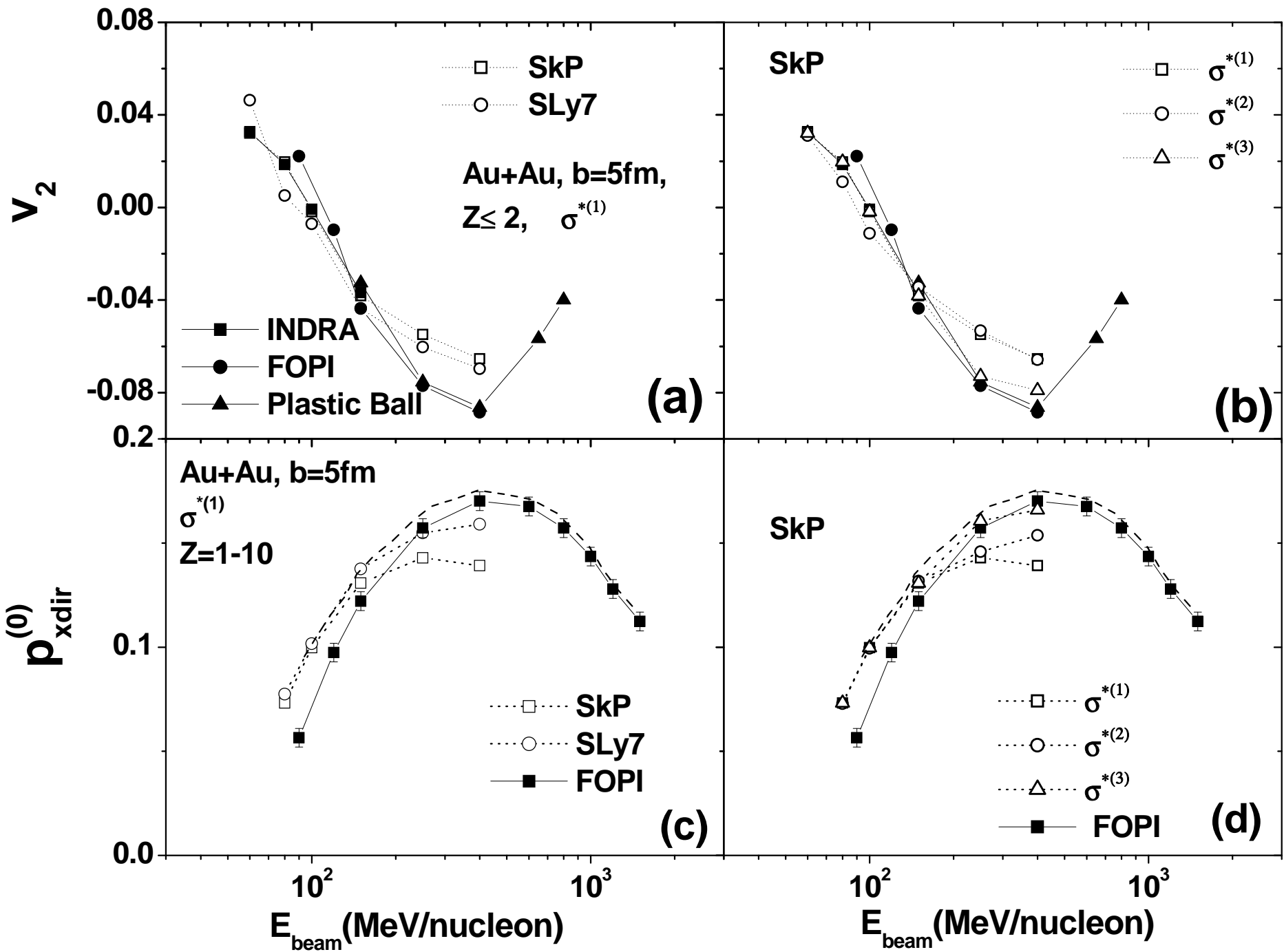
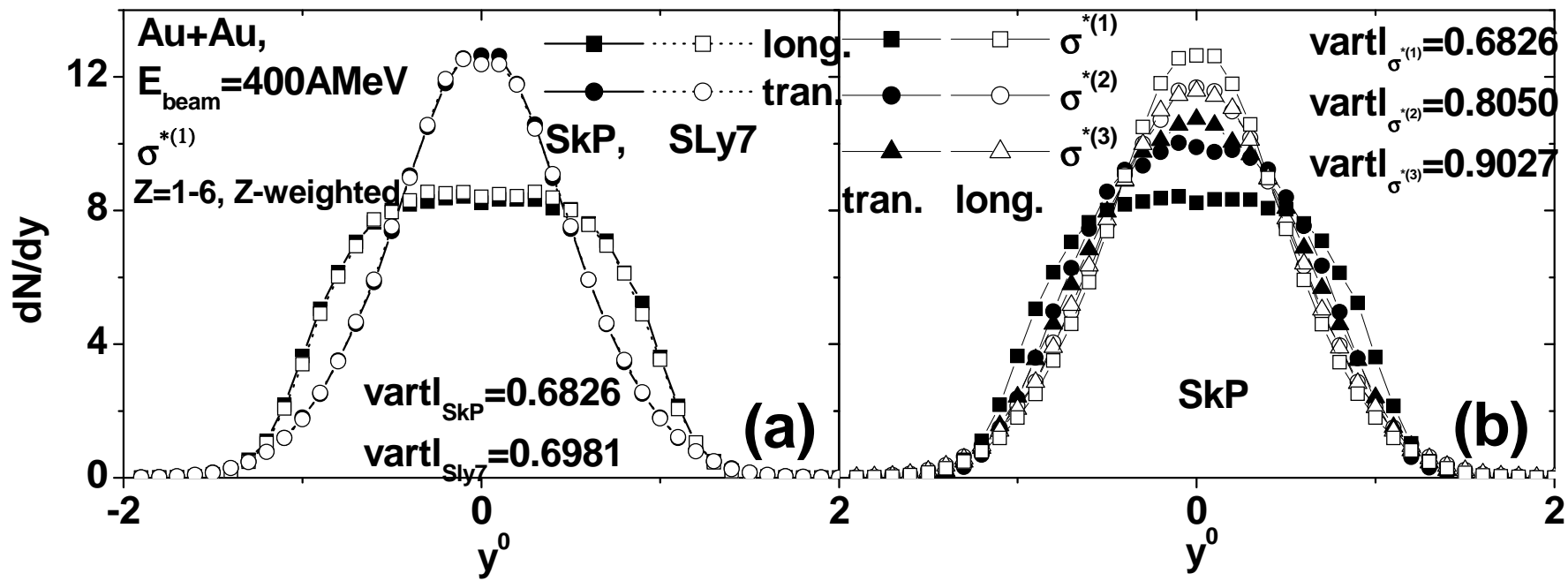
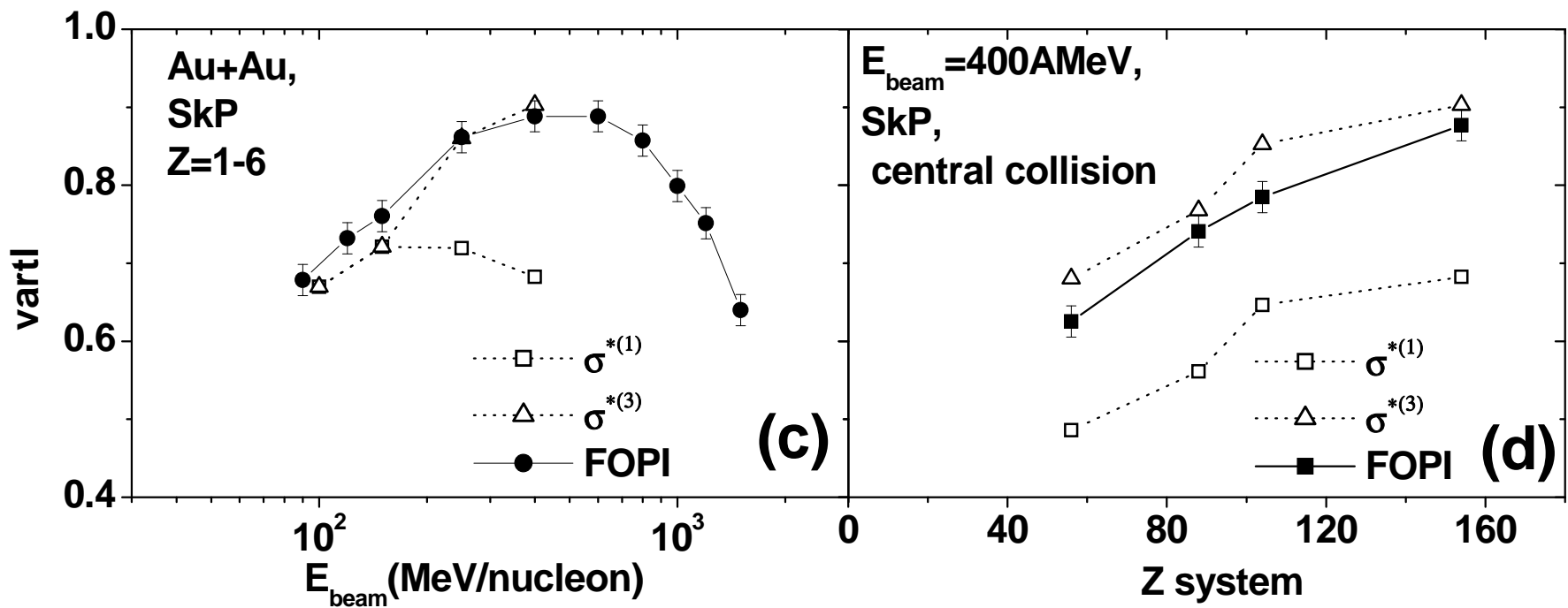
\end{document}